\documentclass[aps,prb, pre,superscriptaddress,twocolumn,final,showpacs,showkeys,nobibnotes,titlepage,twoside,10pt]{revtex4-1}
\usepackage{amsmath}
\usepackage{graphicx}
\usepackage{amsfonts}
\usepackage{amssymb}
\usepackage{subfigure}
\usepackage{float}
\usepackage{color,soul}
\usepackage{natbib}
\usepackage{bm}
\usepackage[english]{babel}

\begin{document}
\raggedbottom
\renewcommand{\figurename}{FIG.}

\author{J. Krausser}
\affiliation{Statistical Physics Group, Department of Chemical Engineering and Biotechnology,
		University of Cambridge, New Museums Site, Pembroke Street, CB2 3RA Cambridge, U.K.}
\author{A. E. Lagogianni}
\affiliation{I. Physikalisches Institut, Universit\"at G\"ottingen, Friedrich-Hund-Platz 1, 37077 G\"ottingen, Germany}
\author{K. Samwer}
\affiliation{I. Physikalisches Institut, Universit\"at G\"ottingen, Friedrich-Hund-Platz 1, 37077 G\"ottingen, Germany}
\author{ A. Zaccone}
\affiliation{Statistical Physics Group, Department of Chemical Engineering and Biotechnology,
		University of Cambridge, New Museums Site, Pembroke Street, CB2 3RA Cambridge, U.K.}
\affiliation{Cavendish Laboratory, University of Cambridge, CB3 0HE Cambridge, U.K.}

\date{\today}

\begin{abstract}
Within the shoving model of the glass transition, the relaxation time and the viscosity are related to the local cage rigidity. This approach can be extended down to the atomic-level in terms of the interatomic interaction, or potential of mean-force. We applied this approach to both real metallic glass-formers and model Lennard-Jones glasses. The main outcome of this analysis is that in metallic glasses the thermal expansion contribution is mostly independent of composition and is uncorrelated with the interatomic repulsion: as a consequence, the fragility increases upon increasing the interatomic repulsion steepness. In the Lennard-Jones glasses, the scenario is opposite: thermal expansion and interatomic repulsion contributions are strongly correlated, 
and the fragility decreases upon increasing the repulsion steepness. 
This framework allows one to tell apart systems where "soft atoms make strong glasses" from those where, instead, "soft atoms make fragile glasses". Hence, it opens up the way for the rational, atomistic tuning of the fragility and viscosity of widely different glass-forming materials all the way from strong to fragile. 

\end{abstract}

\title{Disentangling
	interatomic repulsion and anharmonicity in the viscosity and fragility of
	glasses}
\maketitle 

\section{Introduction}
One of the most puzzling properties of glass-forming liquids is the huge increase of viscosity, by many orders of magnitude, within a narrow range of temperature $T$ upon approaching the glass transition temperature $T_{g}$. As a consequence, considerable interest is being devoted to understanding this phenomenon in terms of the underlying atomic-level structure and dynamics of supercooled liquids~\cite{Debenedetti}. 

For many decades, practically all experimental measurements of viscosity or relaxation time as a function of $T$ upon cooling towards the glass transition $T_{g}$, have been fitted with the Vogel-Fulcher-Tammann (VFT) relation~\cite{Debenedetti_book}. This empirical formula, with three fitting parameters, can capture the exponential or faster-than-exponential increase of viscosity on cooling of practically all cooperative liquids, something that physical theories have struggled to predict on a microscopic basis. Mode-coupling theory predicts a power-law increase of viscosity, whereas entropy-based theories such as Adam-Gibbs and random-first-order are relatively successful in capturing the exponential \cite{Loidl2016,Berthier} trend using activation concepts. 
They involve growing length-scales and moreover can produce predictions of the different steepness of viscosity as a function of temperature~\cite{Mauro2009}. An extended temperature-volume version of this model was formulated~\cite{Masiewicz2012}, which predicts the fragility parameter $m$
for systems with different composition.

As shown by Angell~\cite{Angell}, while some systems (strong glass formers) have a viscosity which follows an Arrhenius $\exp (a/T)$ dependence on $T$, other systems (fragile glass formers) have a much steeper super-exponential dependence on $T$. Various attempts have been reported to formulate physical models that can explain this fundamental observation in terms of the underlying microstructure, dynamics and interaction picture. \\

In spite of these efforts, the VFT relation still lacks a microscopic derivation and most available closed-form expression for the viscosity are semi-empirical extensions of VFT where some parameter is given a tentative physical meaning in terms of a dynamical or structural parameter~\cite{Kelton}.\\ 
Recently, a different approach has been proposed~\cite{Krausser,LagogianniJSTAT2016}, which combines the shoving model of the glass transition~\cite{Dyre2006,Johnson2005} with the atomic theory of elasticity~\cite{Zaccone2011a,Zaccone2011b}. 

There exists experimental support for both the shoving model~\cite{Hecksher2015} and the Adams-Gibbs conjecture~\cite{Bauer2013}. The contributions of the Adam-Gibbs mechanism is, however, not enough to explain the behavior of the fragility, it has to be complemented with the effect of anharmonicity~\cite{Buchenau2014}. Applying the shoving model and the adding the anharmonic effects to describe the fragility of glass-forming liquids is the focus of the present approach.

In the shoving model of the glass transition~\cite{Dyre1998,Dyre2006}, the relaxation time is an Arrhenius function of $T$ where the activation energy is provided by the local rigidity of the cage. In other words, it is the energy needed to break and mobilize the cage for a particle to escape, according to the original idea of Eyring~\cite{and1936}. Hence, $\eta \sim \exp(G/k_{B}T)$ \cite{Johnson2007}, where $G$ is the high-frequency shear modulus which describes the rigidity of the glassy cage. 
Within this approach it is possible to make a step further and use the atomic theory of elasticity to relate the shear modulus $G$ to the average number $Z$ of mechanically-active interatomic connections per atom, $G\propto Z$ for the high-frequency affine modulus. Clearly, if $G$ and $Z$ are relatively insensitive to any change in $T$, the relaxation time and the viscosity are a simple Arrhenius function of $T$, which works well for strong glasses. In the opposite scenario, $Z$ may be a strong function of $T$ because of thermal expansion: if the attractive part of the interatomic potential is shallow, atoms leaving the cage encounter little resistance, hence the viscosity drops dramatically upon increasing $T$ and one recovers the fragile-glass limit.

\section {Disentanglement of the interatomic potential in viscosity and fragility }
Here we apply these ideas to two key systems: real metallic glasses on one hand, and Lennard-Jones model glasses, on the other.
The analysis of these systems is much revealing: it is possible to fully disentangle the contributions of different segments of the interatomic potential (repulsive and attractive) to the viscosity and fragility. This outcome is of utmost importance for developing rational guidelines in the design of glassy materials with tunable mechanical and viscoelastic properties. In our recent work~\cite{Krausser}, an analytical model is proposed that describes the elasticity, viscosity and fragility of metallic glasses in relation to their atomic-level structure and the effective interatomic interaction. 
The model, which has only one adjustable parameter (the characteristic atomic volume for high-frequency cage deformation) is tested against new experimental data from MD simulations of ZrCu alloys and provides an excellent one-parameter description of the viscosity down to the glass transition temperature.\\
We also consider the widely-used Lennard-Jones (LJ) potential defined as: $V(r)=\frac{\epsilon}{(q-p)}[p(\frac{r_{0}}{r})^{q}-q(\frac{r_{0}}{r})^{p}]$, where $\epsilon$ is the depth of the minimum and $r_{0}$ is the position of the energy minimum along the radial coordinate $r$. In this formulation, the LJ potential is a very versatile model system in which the anharmonicity can be varied by tuning the values of power-law exponents $(q,p)$. LJ potentials with different anharmonicity as given by different values of the pair $(q,p)$ are plotted in Fig.1b.
This potential has been extensively used in numerical simulations starting with the pioneering work of Kob and Andersen to make binary mixtures that undergo glass transition upon decreasing the temperature. 

The influence of attractive forces in LJ systems has been studied in detail in Ref.~\cite{Berthier2009}, where it was concluded that a variation of attractive intermolecular forces have only a small influence of the static structure of the LJ glass, but may drastically alter the dynamical relaxation behavior, i.e. the viscosity or relaxation time.

Anharmonicity in this model system can be quantified in various ways. For example, in Ref.~\cite{Bordat}, anharmonicity was quantified by the radial distance $\xi$ at which $V(\xi)=-0.5$. 
In the older literature, a different measure of anharmonicity is given by the cubic coefficient $\zeta<0$ in the Taylor expansion of the potential about the minimum. Classical arguments by Y. Frenkel show that the linear thermal expansion coefficient is proportional to $\mid\zeta\mid$. \\
 Hence, a direct relationship exists between the thermal expansion coefficient $\alpha_{T}$ and the attractive anharmonic tail of the potential, as quantified by either $\xi$ or $\zeta$. In the same way, a similarly global parameter $\lambda$ represents the effect of the repulsive part of the potential, as depicted in Fig.1 (a) and is directly related to the short-range ascending slope of the
 radial distribution function $g(r)$ through the following
 simple power-law expression~\cite{Krausser,LagogianniJSTAT2016}: 
 \begin{equation}
 g(r)\sim(r-\sigma)^{\lambda} 
 \end{equation}
 where $\sigma$ stands for the ion core diameter. 
  
The complex relationship between these two interaction parameters and its impact on viscosity and fragility is explored and disentangled in the following.

\section{ $\alpha_{T} T_{g}$ versus  $\lambda$ for metallic and LJ glass-formers}
We study the relation between $\lambda$ and $\alpha_{T} T_{g}$ for a number of metallic glass-formers, on one hand, and for LJ systems with varying power-law exponent pairs, on the other, in a comparative framework. 

The proposed interatomic potential of our recent work has been applied to experimental data of metallic alloys (see Appendix A). In some cases, $\lambda$ could be extracted from experimental data of $g(r)$, whereas in other cases it has been fitted to the viscosity data. In the fitting of metallic glass data using our proposed interatomic potential, the values of $T_{g}$ and $\alpha_{T}$ were taken from the literature. Hence, using those fittings, it is possible to analyse the interrelation between  $\lambda$ and $\alpha_{T} T_{g}$, and the impact thereof on the fragility $m$.

For the model LJ systems, instead, we use the simulation data of Bordat et al.~\cite{Bordat} from the literature who studied three different LJ systems with different values of the power-law exponents as depicted in the Fig.1 (b). 

	\begin{figure}
		\centering
		\includegraphics[width=0.94\columnwidth]{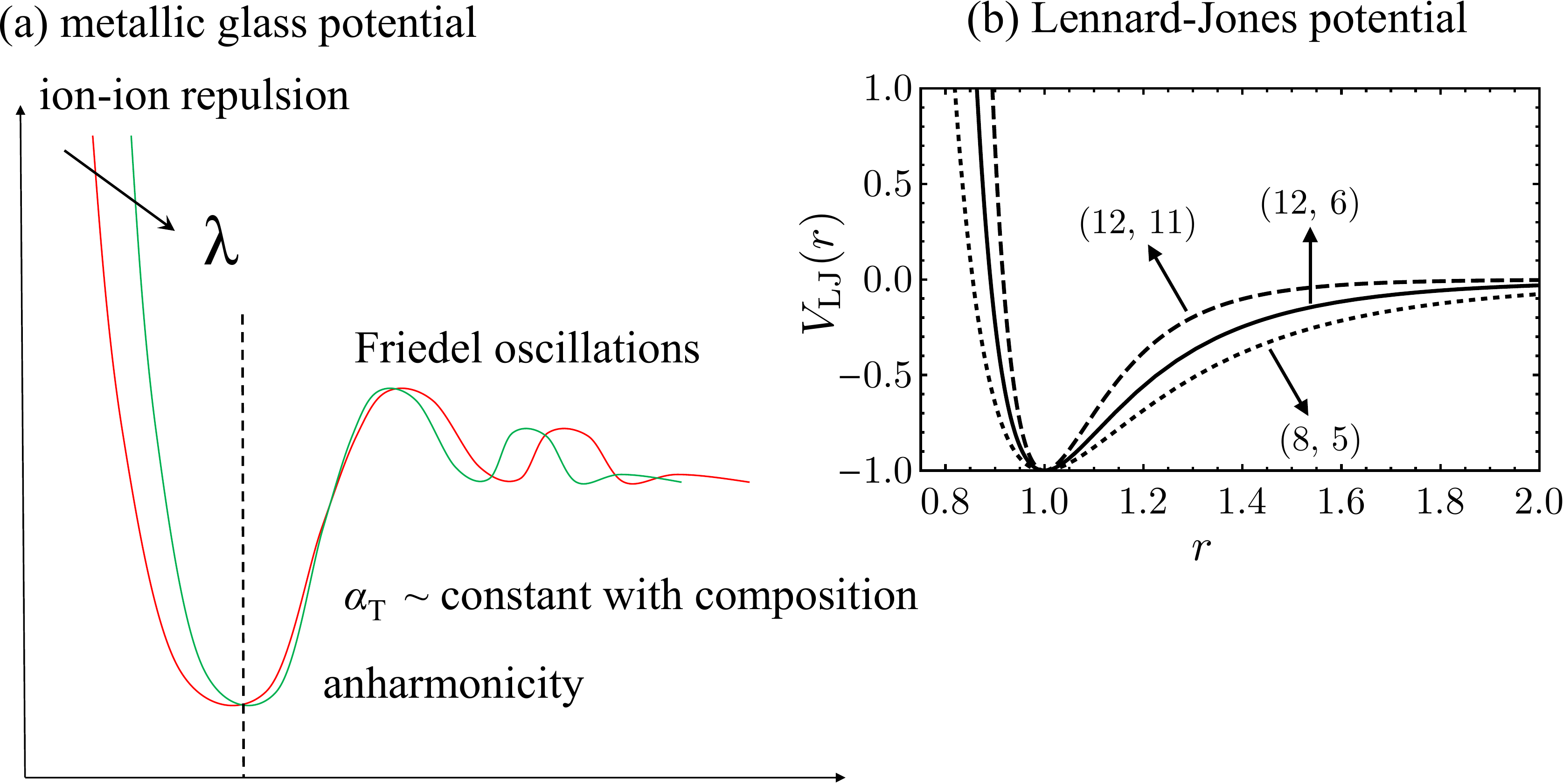}
		\caption{Ashcroft-Born-Mayer interatomic potential.
					Representation of the Ashcroft–Born–Mayer interatomic  (see Appendix A) using the one-parameter logarithmic expression in terms of the global interaction parameter $\lambda$ (including the two separate contributions to the interatomic potential). This illustrative plot was generated for a repulsive steepness $\lambda =99.7$. 
		}
		\label{fig:rn}
	\end{figure}
	
The $\alpha$-relaxation time $\tau_{\alpha}$ was measured in Ref.~\cite{Bordat}, as a function of the reduced temperature $T/T_{ref}$ for the three LJ systems, where $T_{ref}\simeq T_{g}$. Upon using that $\tau \propto \eta$, we fitted the simulation data for $\tau$ using our interatomic potential (see Appendix A). The proportionality constant between $\tau$ and $\eta$ is absorbed in the parameter $\eta_{0}$. Furthermore, to keep things simple, we also used the product $V_c C_G$, where $V_c$ corresponds to the characteristic atomic volume and $C_G$ to the shear modulus value at the glass transition temperature $T_g$ (see Appendix A), as the only fitting parameter. 
The values of $\lambda$ are determined by fitting the ascending flank of the first peak of $g(r)$ (for the majority particle species, since it is a binary mixtures) as reported in Bordat et al.~\cite{Bordat}, according to the procedure reported in Ref.~\cite{LagogianniJSTAT2016}. 
The results are shown in Fig. 2. 

\begin{figure}
\centering
\includegraphics[width=0.94\columnwidth]{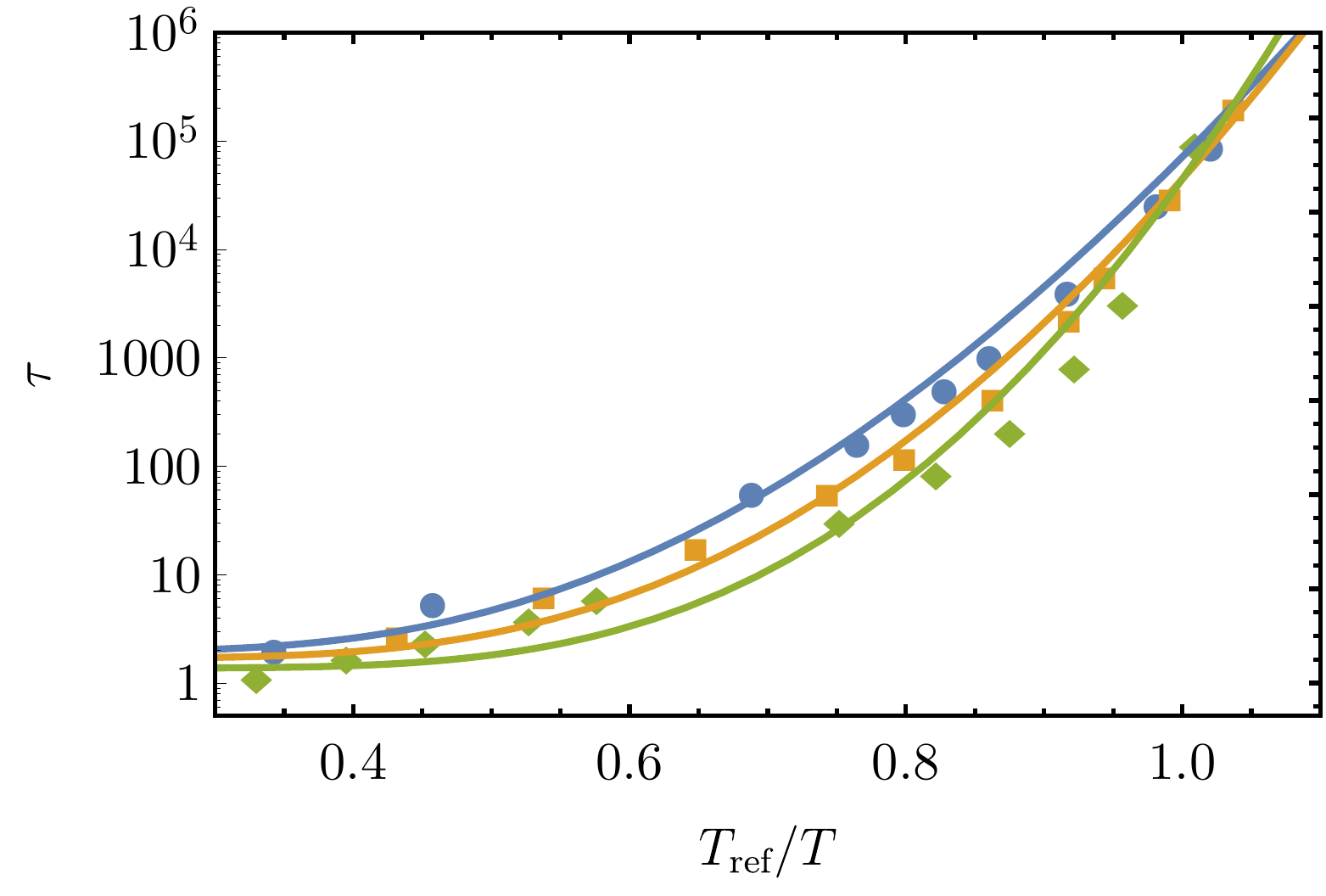}
\caption{
Theoritical fittings of simulation data of the LJ systems.
Symbols are simulation data for the $\alpha$-relaxation time measured in the simulations of the model LJ systems with variable exponents in Bordat et al. Solid lines are theoretical fittings using our effective interatomic potential (see Appendix A).  
}
\label{figure:rn}
\end{figure}

\begin{figure} 
	{\includegraphics[width=0.95\columnwidth]{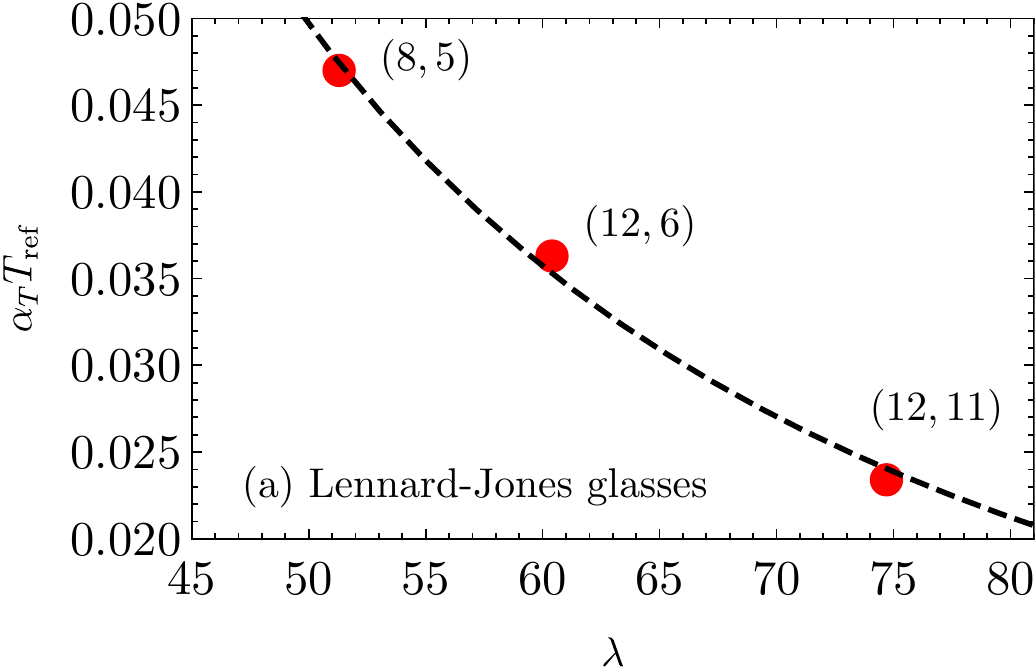}}
	{\includegraphics[width=0.95\columnwidth]{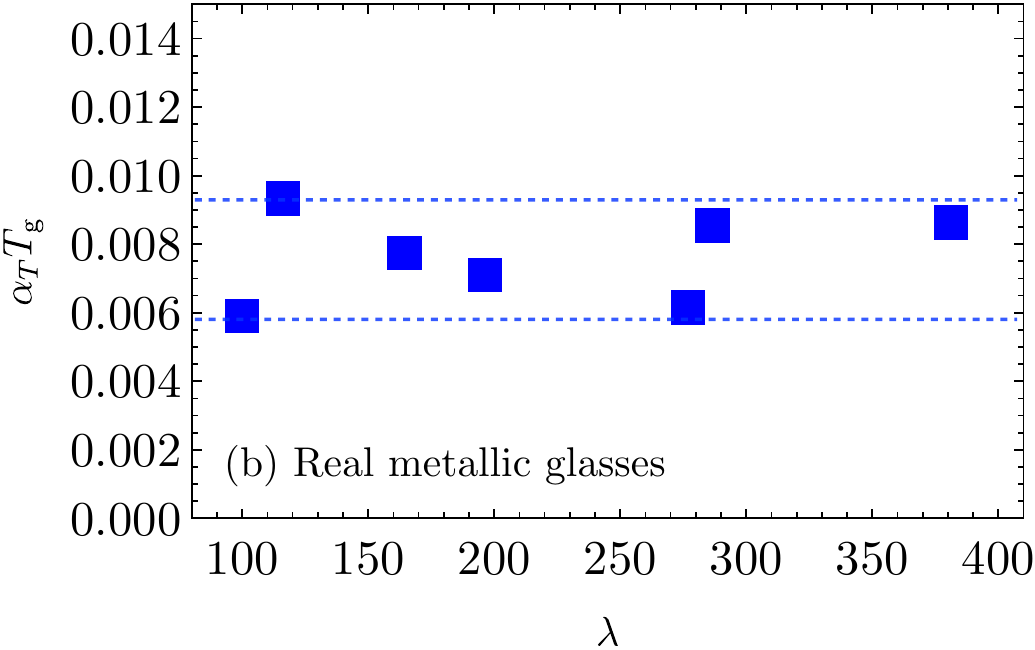}}
	{\includegraphics[width=0.95\columnwidth]{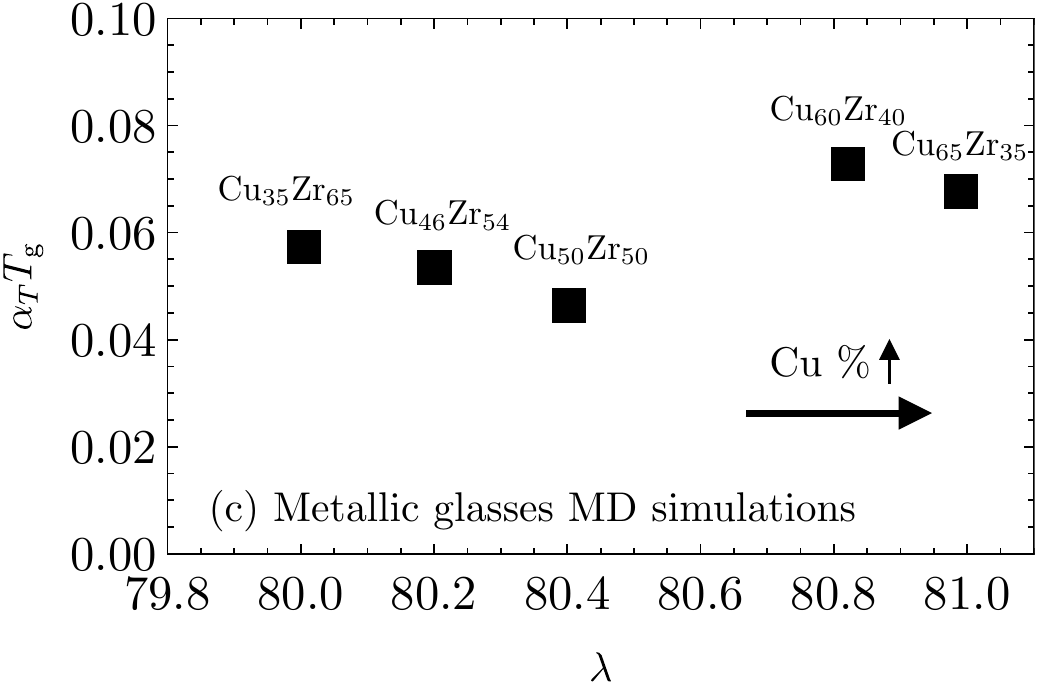}}
\caption{$\alpha_{T} T_{g}$ versus $\lambda$-Experiment and simulations.
(a) Fitted values of $\alpha_{T} T_{ref}$ for the three different LJ systems as a function of the corresponding value of interparticle repulsion $\lambda$.  (b) literature values of $\alpha_{T} T_{g}$ ($\alpha_{T}$ is the linear thermal expansion coefficient) of various metallic glasses as a function of the corresponding value of $\lambda$ as fitted using Eq.(1) in Ref.~\cite{Krausser}. From left to right (increasing value of $\lambda$): Zr$_{46.75}$ Ti$_{8.25}$ Ni$_{10}$ Cu$_{7.5}$ Be$_{27.5}$, Pd$_{43}$ Cu$_{27}$ Ni$_{10}$ P$_{20}$, Pt$_{57.5}$ Ni$_{5.3}$ Cu$_{14.7}$ P$_{22.5}$, La$_{55}$ Al$_{25}$ Ni$_{20}$, Zr$_{41.2}$ Ti$_{13.8}$ Ni$_{10}$ Cu$_{12.5}$ Be$_{22.5}$, Pd$_{40}$ Ni$_{40}$ P$_{20}$,  Pd$_{77.5}$ Cu$_{6}$ Si$_{16.5}$.
(c) values of $\alpha_{T} T_{g}$ ($\alpha_{T}$ is the volumetric thermal expansion coefficient) of the ZrCu metallic glass extracted from MD simulations for different stoichiometries as a function of $\lambda$ calculated from the total $g(r)$ of the corresponding system by the way described in our recent work\cite{LagogianniJSTAT2016} (Appendix A)}
\end{figure}

\begin{figure} 
	{\includegraphics[width=0.95\columnwidth]{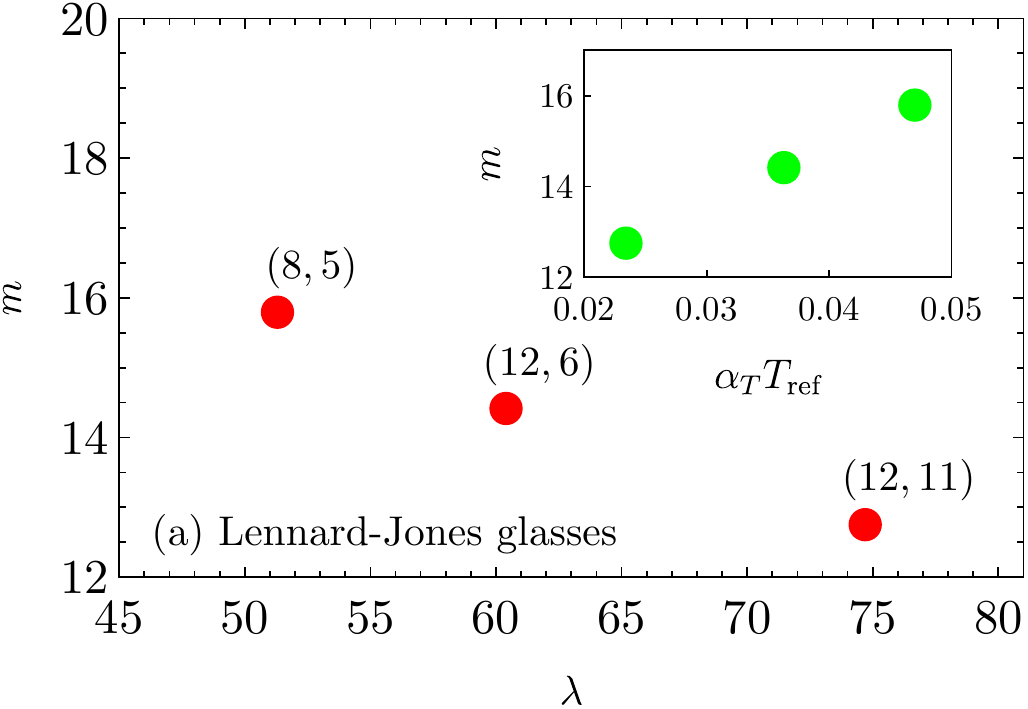}}
	{\includegraphics[width=0.95\columnwidth]{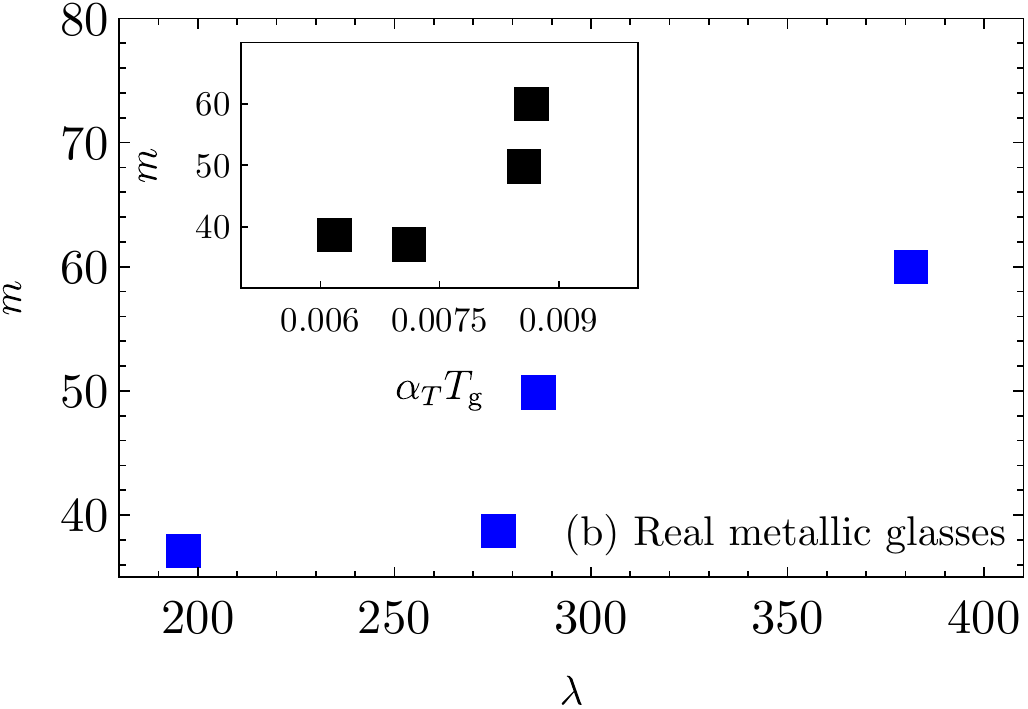}}
\caption{Fragility $m$ as a function of interatomic repulsion steepness $\lambda$.
(a) for the LJ systems, while the inset shows how the fragility $m$ depends on the thermal expansion parameter $\alpha_{T} T_{ref}$.  (b) for various metallic glasses: from left to right (increasing value of $\lambda$):
 La$_{55}$ Al$_{25}$ Ni$_{20}$, 
Zr$_{41.2}$ Ti$_{13.8}$  Ni$_{10}$ Cu$_{12.5}$ Be$_{22.5}$, 
Pd$_{40}$ Ni$_{40}$ P$_{20}$, Pd$_{77.5}$ Cu$_{6}$ Si$_{16.5}$. The inset shows how the fragility $m$ depends on the thermal expansion parameter $\alpha_{T} T_{g}$.}

\end{figure}

From the fittings, we thus obtain the values of $\alpha_{T}$ for the three LJ systems. As expected based on lattice dynamical considerations, $\alpha_{T}$ increases with increasing the anharmonicity of the LJ potential, i.e. from the most harmonic-like to the most anharmonic, in the following order: $(12,11) \rightarrow (12,6)\rightarrow (8,5)$. Also, we plotted the values of the product $\alpha_{T} T_{ref}$ as a function of $\lambda$ in Fig. 3(a), which enter the expression for the fragility $m$ as a product. By virtue of the definition of the LJ potential, $\alpha_{T}$ is a decreasing function of  $\lambda$, which is also evident from the plots in Fig.~1(a). %Moreover, this behavior is confirmed by fitting the double-exponential relaxation time function to the data available in Ref.~\cite{Bordat}.
In addition to that, the product $\alpha_T T_{ref}$ follows an empirical power-law $\alpha_{T} T_{ref} \sim \lambda^{-1.79}$. Upon inserting this into the expression for $m$, we thus obtain the scaling relation $m \sim \lambda^{-0.79}$. This implies that, for model LJ systems, a softer interatomic repulsion is linked to more fragile glass-former. This is the opposite to what is observed in soft repulsive colloids and in metallic glasses, where "soft atoms/particles make strong glass-formers". 
It is important to emphasize that in the LJ systems studied, the increase in anharmonicity in the attractive part of the potential (which sets the thermal expansion coefficient $\alpha_{T}$) goes hand in hand with softer short-ranged repulsion by construction. This is indeed reflected in the fact that $\alpha_{T}$ is a decreasing function of $\lambda$.

Let us now consider the situation with metallic glasses. In Fig.~3(b), we have plotted values of $\alpha_{T} T_{g}$ ($\alpha_{T}$ is the linear thermal expansion coefficient) of various real metallic glasses taken from the experimental literature, as a function of the corresponding $\lambda$ values obtained in Ref.\cite{Krausser} by fitting experimental viscosity data with Eq. (5) (see Appendix). The experimental data sets were obtained at atmospheric pressure conditions~\cite{Johnson2007}. Additionally, in Fig. 3(c), the $\alpha_{T} T_{g}$ ($\alpha_{T}$ is the volumetric thermal expansion coefficient) values versus $\lambda$ are plotted for a given metallic system, studied by MD simulations, where the role of the stoichiometry has been added. The situation in both cases looks as follows: there is no appreciable trend and $\alpha_{T} T_{g}$ appears to be muss less dependent on $\lambda$ as in the LJ case. In particular, $\alpha_{T} T_{g}$ varies very little over a comparatively much larger $\lambda$-interval, within a band of $\pm 0.00131 $ of its average value $0.00764$, in the case of the real metallic glasses. This behavior has been observed experimentally for a wide range of different bulk metallic glasses~\cite{Kato2008}. In the second case of the CuZr binary alloy of varying composition it varies as $\pm 0.00913 $ around its average value $0.06266$.

In the case of LJ, instead, we had not only a clear decreasing trend, but also a larger variation in 
$\alpha_{T} T_{ref}$, by more than a factor of $2$, over a comparatively much narrower $\lambda$ interval.
Hence, we can set approximately $\alpha_{T} T_{g}\sim \lambda^{0} \sim const$ in Eq.~(6) (see Appendix) for the case of metallic glasses, which gives $m \sim \lambda$ for the fragility. This clearly explains that "soft atoms make strong glasses" in the case of metallic glasses, because the product of the thermal expansion coefficient and the glass transition temperature in this case is practically independent of the interatomic repulsion. This is a very subtle but important point and a deep insight into our understanding of glasses.

Let us now consider what are the consequences for the fragility, by looking at the plots of $m$ as a function of $\lambda$ and as a function of $\alpha_{T}T_{ref}$ and  $\alpha_T T_g$ for the two classes of systems, respectively. 
In Fig. 4(a), main panel, we plotted the fragility as a function of $\lambda$ for the LJ systems. As discussed above, in this case $m$ is a decreasing function of $\lambda$,  even though $m$ increases with increasing $\alpha_{T}T_{ref}$, because $\alpha_{T} T_{ref}$ is a nonlinearly decreasing function of $\lambda$. Hence, according to our expression for the fragility, $m=\frac{1}{\ln 10}\frac{V_{c} C_{G}}{k_{B}T_{g}}[1+(2+\lambda)\alpha_{T} T_{g}]$, derived in the
Appendix, the overall dependence of $m$ is that it decreases in the order $(8,5) \rightarrow (12,6)\rightarrow (12,11) $. 

In Fig. 4(b), main panel, we plotted the fragility as a function of $\lambda$ for the metallic glasses. As anticipated, here we have the opposite trend: "soft atoms make strong glasses", and the fragility increases upon increasing the repulsion steepness $\lambda$. However, let us look also at the dependence of $m$ on $\alpha_{T}T_g$: it appears to be qualitatively the same dependence seen for the LJ systems, i.e. the fragility increases with increasing the product of the thermal expansion coefficient and the glass transition temperature, $\alpha_{T} T_g$, but the trend is comparatively less pronounced for the same reasons mentioned in the context of Fig. 3(b).
\section {Conclusions}

Hence, we have shown that the fragility of the glass-formers investigated here under isobaric conditions is an increasing function of the product of the thermal expansion coefficient and the glass transition temperature, which is directly related to the \textit{attractive} anharmonicity for the model LJ glass-formers.
The situation, however, is different for the repulsion steepness (and its inverse, the softness). In LJ systems, by construction, the repulsion steepness decreases upon increasing the attractive anharmonicity, and the fragility is a decreasing function or the repulsion steepness parameter $\lambda$ because $\alpha_{T}T_\text{ref}$ is a stronger-than-linear decreasing function of $\lambda$ in the fragility formula (Eq.(6) Appendix A). Hence, the overall dependence of the fragility $m$ on the repulsion steepness is strongly influenced by the strong dependence of $\alpha_{T}T_{ref}$ on $\lambda$.
For metallic glasses, the situation is reverse: $\alpha_{T}T_g$ is basically independent of $\lambda$. Hence the overall dependence of $m$ on $\lambda$ in this case is of direct proportionality as given by the expression for $m$ quoted above (Eq. (6) in the Appendix).

In physical terms, this analysis clarifies that an increased fragility is associated with larger values of the product $\alpha_T T_g$, which implies larger atomic mobilities at the level of third and 4th neighbours~\cite{Busch}, and the ability of the system to rearrange into more stable local configurations upon decreasing $T$. At the same time, however, the fragility also increases upon increasing the short-range repulsion steepness, because this implies a steeper $T$-dependence of the local cage rigidity encoded in the high-frequency shear modulus $G$. This, in turn, is controlled by the local coordination number $Z$ and hence by the $T$-dependence of $Z$~\cite{Zaccone2013}, which is a function of the repulsion steepness $\lambda$. The conceptual framework developed here thus mechanistically explains the apparent contradiction in the recent literature (that "soft atoms make strong glasses" for metals~\cite{Krausser} and colloids~\cite{Weitz}, whereas "soft atoms make fragile glasses", for LJ systems~\cite{Bordat}) in terms of the underlying interaction physics. This framework will prove useful to achieve a rational design of mechanical properties of metallic glasses and other amorphous advanced materials. 

The support of the EU through VitrimetTech ITN network FP7-PEOPLE-2013-ITN-607080 and FG 1394P1 is thankfully acknowledged.

%\makeatletter
%\renewcommand\@biblabel[1]{#1.}
%\makeatother

\section{APPENDIX A. Effective Ashcroft-Born-Mayer pseudopoential-Analytical expressions for the high frequency shear modulus and the viscosity}	
In recent work~\cite{Krausser}, we analysed several amorphous metallic alloys in an attempt to extract an effective, averaged interatomic potential which describes the short-range repulsion between any two ions in a metallic alloy melt. Based on the systematic fitting of shear modulus and viscosity data for various three- and 5-component alloys we proposed the following interatomic potential which comprises two contributions: (i) the longer-ranged Thomas-Fermi (screened-Coulomb) repulsion modulated by the Ashcroft correction and (ii) the Born-Mayer closed-shell repulsion due essentially to Pauli repulsion. 
\begin{equation}
V(r)=A \frac{\exp^{-q_{\text{TF}}(r-2a_0)}}{r-2a_0}+B\mathrm{e}^{ -C(r-\bar{\sigma})}
\end{equation}
where 
\begin{equation}	A=Z_{\text{ion}}^2 e^2\cosh^{2}(q_{\text{TF}}  R_{\text{core}})
\end{equation}

The Thomas-Fermi contribution is more long-ranged and is described by a Yukawa-potential type expression. The Born-Mayer contribution is a simple exponentially-decaying function of the core-core separation, motivated by the radial decay of electron wavefunctions for the closed shells. The effective (average) interatomic potential for two atoms in a metallic glass or melt is schematically depicted in Fig.1 of the main article. 

We also consider the standard Lennard-Jones potential defined as: $V(r)=\frac{\epsilon}{(q-p)}[p(\frac{r_{0}}{r})^{q}-q(\frac{r_{0}}{r})^{p}]$, where $\epsilon$ is the depth of the minimum and $r_{0}$ is the position of the energy minimum along the radial coordinate $r$. Anharmonicity in this model system can be quantified in various ways. For example, in Ref.~\cite{Bordat}, anharmonicity was quantified by the radial distance $\xi$ at which $V(\xi)=-0.5$. 
In the old literature, a different measure of anharmonicity is given by the cubic coefficient $\zeta<0$ in the Taylor expansion of the potential about the minimum. Classical arguments by Y. Frenkel show that the linear thermal expansion coefficient is proportional to $\mid\zeta\mid$.

Hence, a direct relationship exists between the thermal expansion coefficient $\alpha_{T}$ and the attractive anharmonic tail of the potential, as quantified by either $\xi$ or $\zeta$. The next step, is to find a similarly global parameter like $\alpha_{T}$ to represent the effect of the repulsive part of the potential. 
This can achieved as follows.

A simple parametrization of the short-range repulsive part of the interatomic potential is obtained by fitting the repulsive ascending part of the radial distribution function $g(r)$ to a power-law: $g(r)\sim (r-\sigma)^{\lambda}$, where $\sigma$ corresponds to the soft-core diameter of the atoms. This fitting is valid between $r\approx0$ and $r\approx r_{0}$, where we approximate the maximum of the first peak of $g(r)$ with the minimum $r_{0}$ of the pair potential. 
Upon inverting the Boltzmann relation $g(r)=\exp(-V_{\mathrm{eff}}(r)/k_{B}T)$, the potential of mean-force $V_{\mathrm{eff}}$ is obtained directly from $g(r)$. The potential of mean-force $V_{\mathrm{eff}}$ reduces to the pair potential $V(r)$ only in the limit of zero density of particles (ideal gas limit). At the high density of supercooled liquids, $V_{\mathrm{eff}}$ crucially contains many-body effects and represents the effective interaction between two particles mediated by the motions of all other particles in the liquid~\cite{Hafner}. At short range, however, $V_{\mathrm{eff}}$ and $V(r)$ are very similar and both dominated by the repulsive part of the interaction (they both diverge as $r\rightarrow 0$). 
Since we are looking for a global repulsion parameter, analogous to $\alpha_{T}$ for the attraction, it is important to work with $V_{\mathrm{eff}}$ rather than $V(r)$. 
	
Hence, we can obtain an estimate of the repulsive part of the potential of mean force using 
\begin{equation}	
V_{\mathrm{eff}}=-\lambda \ln(r-\sigma).	
\end{equation}
where $\lambda$ 
comes from a power-law fitting of $g(r)$ up to the maximum of the first peak as described above and in Ref.~\cite{Krausser,LagogianniJSTAT2016}.

The next step, is to find a way to connect the shear modulus $G$, which enters the shoving model, to the global interaction parameters, $\lambda$ and $\alpha_{T}$. 
The high-frequency (affine) shear modulus can be written using lattice dynamics as $G = \frac{1}{5 \pi} \frac{\kappa}{R_0} \phi Z$, where $R_{0}$ is some average (snapshot) distance between two nearest-neighbour atoms in the equilibrated supercooled liquid, $\kappa$ is the harmonic spring constant (i.e. the curvature of the main energy minimum in Fig.1), and $\phi$ is the atomic packing fraction. 
	
Using the fact that the upper integration limit of $r_{max}$ increases with the packing fraction $\phi$, integrating the $g(r)$ up to a threshold which is proportional to $\phi$, as done in Ref.~\cite{LagogianniJSTAT2016}, yields the scaling law $Z \sim  \phi ^{1+\lambda}$. Although the upper limit of the integral could be perhaps identified with $r_{\mathit{max}}$, since we are interested here in the qualitative behaviour we prefer to leave it as a generic threshold $\propto\phi$ such that the limit $Z\rightarrow 0$ is correctly recovered when $\phi \rightarrow 0$.
	
Moreover, the definition of the Debye-Grueneisen thermal expansion coefficient $\alpha_{T}$, in terms of the atomic packing fraction $\phi = vN/V$ (with $v$ the characteristic atomic volume and $N$ the total number of ions in the material) gives $\phi(T) \sim \mathrm{e}^{-\alpha_T T}$, as discussed e.g. in Ref.~\cite{Zaccone2013}. According to this result, $\phi$ decreases with increasing temperature $T$, an effect mediated by the thermal expansion coefficient defined as $\alpha_T = \frac{1}{V}(\partial V/ \partial T) = -\frac{1}{\phi}( \partial \phi/ \partial T)$.
	
Replacing the latter relationship between $\phi$ and $T$ in the expression for $Z$,  we finally obtain a closed-form equation which relates $ G $ to the two global interaction parameters, the short-range repulsion parameter $\lambda$ and the attraction anharmonicity parameter $\alpha_{T}$, 
$G(T) =\dfrac{1}{5 \pi} \dfrac{\kappa}{R_0}\exp[ -(2+\lambda) \alpha_T T ]$.
	
	%Equation (1) accounts, in compact form, for all the salient features of the interatomic interaction, and contains the effect of repulsion steepness (short-ranged part of $V(r)$) as expressed by $\lambda$, and of anharmonicity, expressed by $\alpha_{T}$. A schematic depiction of how the global parameters $\lambda$ and $\alpha_{T}$ are related to features of $V(r)$, is presented in Fig. 2. It is important to note that the longer-ranged, anharmonic attractive part of the interaction in metals also stems from non-local, volume-dependent terms in the interaction of the ions with the partly delocalized electron gas. 
	%Hence, a microscopic description in terms of pair-interactions alone is generally not valid, although for volume-preserving shear deformations, as considered here, it can still be used. 
	
The above expression for $G$ can be rewritten as
\begin{align}\label{eq_th_shear}
G(T)=C_G\exp{\left[	\alpha_T T_g (2+\lambda)\left(1-\frac{T}{T_g}\right)\right]}.
\end{align}
where $C_G=\frac{\varepsilon}{5\pi}\frac{\kappa}{R_0}\mathrm{e}^{-\alpha_T T_g (2+\lambda)} $ is defined as the shear modulus value at the glass transition temperature $T_{g}$, i.e. $C_{G}\equiv G(T_{g})$. The constant $\varepsilon$ stems from the integration of $\alpha_T$ and from the dimensional prefactor in the power-law ansatz for $g(r)$. All the parameters in this expression are either fixed by the experimental/simulation protocol or can be found in the literature. The parameter $\lambda$ has to be extracted from $g(r)$ data, according to the protocol that we give in the Section VI of Ref.~\cite{LagogianniJSTAT2016}.
	
We can now use our model for $ G(T) $ to evaluate the activation energy $E(T)$ involved in restructuring the glassy cage and, hence, the viscosity $\eta(T)$ of the melts. Within the framework of the cooperative shear or shoving model of the glass transition~\cite{and1936, Dyre1998, Dyre2006}, the activation energy for local cooperative rearrangements is $E(T)=G V_{\text{c}}$. The characteristic atomic volume $V_{\text{c}}$ showing up here is accessible through the theoretical fitting of the viscosity data, although its value cannot be arbitrary and it must be representative of the atomic composition of the alloy and of the atomic sizes of its constituents~\cite{Krausser,LagogianniJSTAT2016}.
Replacing the expression for $E(T) $ in the Arrhenius relation given by the cooperative shear model of the glass transition, and using Eq.~\eqref{eq_th_shear} for $ G(T) $ inside $E(T)$, we obtain the following analytical expression for the viscosity, 
\begin{align}\label{thvisc}
\dfrac{\eta(T)}{\eta_{0}}=\exp{\left\{ \dfrac{V_c C_G}{k \, T}	\exp{\left[	(2+\lambda)	\alpha_T T_g \left(	1-\frac{T}{T_g}\right)	\right]}\right\}}.
\end{align}
where $\eta_0$ is a normalisation constant set by the high-$T$ limit of $\eta$.
	
It is important to consider how the \textit{double-exponential} dependence of the viscosity on the temperature arises. The first exponential stems from the elastic activation described in the framework of the cooperative shear model, whereas the second exponential is due to the Debye-Gr\"uneisen thermal expansion rooted in lattice-dynamical considerations of anharmonicity. This formula accounts for both anharmonicity, through $\alpha_{T}$, and for the repulsion steepness $\lambda$ (or softness $1/\lambda$). It is clear that, depending on the mutual inter-relation between $\lambda$ and the thermal expansion factor $\alpha_{T} T_{g}$, the viscosity may be affected in a different way by the different sectors of the interatomic interaction as depicted in Fig.1. For example, in previous work, it was found that $\eta$ is a sensitive function of $\lambda$ meaning that larger values of $\lambda$ are associated with a steeply rising viscosity, and viceversa. This is reflected in the relation for the fragility $m$ which can be readily derived from Eq.(5), and gives 
\begin{equation}
m=\frac{1}{\ln 10}\frac{V_{c} C_{G}}{k_{B}T_{g}}[1+(2+\lambda)\alpha_{T} T_{g}].
\end{equation}

\section {APPENDIX B. Simulation Model}  
We performed Molecular Dynamics (MD) simulations of the $Cu_xZr_{100 - x}$ system (where $x= 35, 46, 50, 60, 65 $),  by employing a semi-empirical many-body potential \cite{duan2005molecular2005} in analogy to the tight-binding scheme in the second-moment approximation~\cite{cleri1993tight1993,rosato1989thermodynamical1989}. The equations of motion were integrated by using the Verlet algorithm with a time step of $5~\mathrm{fs}$. The systems of $1.28\times 10^{5}$ atoms were prepared by equilibrating them at $300~\mathrm{K}$ in NPT ensemble (zero pressure) for $100~\mathrm{ps}$ and subsequently heated up to $2000~\mathrm{K}$ for melting. After their equilibration in the liquid state, the configurations were cooled down to $300~\mathrm{K}$ (NPT) with a cooling rate of $10~\mathrm{K/ps}$, where they were finally equilibrated for $100~\mathrm{ps}$ in a NPT ensemble(zero pressure). In all the production simulation runs the temperature and the pressure were kept constant by coupling the system to a Nose\cite{Nose1984} thermostat and to an Andersen\cite{Andersen1980} barostat respectively. Upon cooling the pressure of the system was maintained zero by alloying the simulation box to change dimensions without changing its shape.

 \setcounter{equation}{0}
\newpage

\end{document}